\documentclass[times]{qjrms4}

\usepackage[colorlinks,bookmarksopen,bookmarksnumbered,citecolor=red,urlcolor=red]{hyperref}
\usepackage{framed}
\usepackage{amssymb}
\usepackage{graphicx}
\usepackage{geometry}
\usepackage{sectsty}
\allsectionsfont{\mdseries}
\usepackage{tgbonum}
\usepackage{caption}
\usepackage{sidecap}
\usepackage{wrapfig}
\usepackage{mathptmx}
\usepackage{tabu}
\RequirePackage{ifthen}
\usepackage{times}
\usepackage{amsmath}
\usepackage{algpseudocode}
\usepackage{enumerate}
\usepackage{multirow}
\usepackage{array}
\usepackage{capt-of}
\usepackage{mathtools}

\DeclarePairedDelimiter\ceil{\lceil}{\rceil}

\newcommand{\BEGINALG}{\begin{algorithmic}[1]}

\DeclarePairedDelimiter\abs{\lvert}{\rvert}

\DeclarePairedDelimiter\floor{\lfloor}{\rfloor}

\newcommand\BibTeX{{\rmfamily B\kern-.05em \textsc{i\kern-.025em b}\kern-.08em
T\kern-.1667em\lower.7ex\hbox{E}\kern-.125emX}}

\usepackage{moreverb}

\def\volumeyear{2013}

\begin{document}

\runningheads{A. Gregory and C. Cotter}{On the Calibration of MLMC Ensemble Forecasts}

\title{On the Calibration of Multilevel Monte Carlo Ensemble Forecasts}

\author{A. Gregory \corrauth and C. J. Cotter}

\address{Department of Mathematics, Imperial College London, UK}

\corraddr{a.gregory14@imperial.ac.uk}

\begin{abstract}

Multilevel Monte Carlo can efficiently compute statistical estimates of discretized random variables, for a given error tolerance. Traditionally, only a certain statistic is computed from a particular implementation of multilevel Monte Carlo. This paper considers the multilevel case when one wants to verify and evaluate a single ensemble that forms an empirical approximation to many different statistics, namely an ensemble forecast. We propose a simple algorithm that, in the univariate case, allows one to derive a statistically consistent single ensemble forecast from the hierarchy of ensembles that are formed during an implementation of multilevel Monte Carlo. This ensemble forecast then allows the entire multilevel hierarchy of ensembles to be evaluated using standard ensemble forecast verification techniques. We demonstrate the case of evaluating the calibration of the forecast in this paper. 

\end{abstract}


\maketitle

\section{Introduction}

Multilevel Monte Carlo (MLMC) \citep{Giles} is a technique that has
gained significant popularity over the past decade. It is designed to
produce statistical estimators for discretized random variables at
significantly lower computational costs than their Monte Carlo
counterparts for a fixed error. This is done by using a hierarchy of larger ensembles
using lower accuracy models, and smaller ensembles using higher
accuracy models. For probabilistic forecasting, one can use this
multilevel technique to estimate statistics from a forecast
probability distribution, given some distribution of the initial conditions
and/or random forcing.

In the multilevel Monte Carlo framework,
one usually considers a particular statistic, such as evaluations of the cumulative distribution function (CDF) \citep{Ritter, Wilson, Elfverson}, probability density function (PDF) \citep{Bierig} or expected values \citep{Giles,Cliffe}, selecting the ensemble sizes / finest level of resolution so that the overall multilevel estimator produces an efficient and
accurate approximation. 

In the case of ensemble forecasting, one
usually wishes to compute many statistics from the same ensemble. 
These approximations can be assessed using suitable verification
techniques. Verification tools used within ensemble forecasting
usually work alongside observations of the process that one is
interested in forecasting and can help verify properties from
calibration to the sharpness of a forecast \citep{Gneiting07}.

Given the multilevel hierarchy of ensembles from different resolutions
that form MLMC estimates of statistics, we would also like to evaluate
/ verify these ensembles in the same way; this is the subject of this
paper. We propose a methodology to take observables of a univariate random variable, or scalar observables of a multidimensional random variable (such as a random field evaluated at a
point in space), from a
multilevel hierarchy of ensembles with varying resolutions and generate an accompanying single
ensemble forecast. Most of the standard techniques in the field of ensemble forecasting are limited to the univariate case; in the context of large dimensional models in weather and climate these are usually applied to scalar observables such as point values or integral quantities.

This single forecast is statistically consistent
with the multilevel estimate.  It can then
be used to verify the forecast from the original ensemble
hierarchy using standard methods such as calibration tests.

An alternative approach to this could be approximating each verification or scoring measure, such as the calibration or sharpness, individually and directly from the multilevel hierarchy of ensembles. For example, one could use a MLMC approximation for the CDF \citep{Ritter, Elfverson} to help compute a rank histogram to evaluate the forecast calibration.
Each different MLMC approximation typically comes with a framework to implement it, such as
a smoothing scheme in the former of those two studies.

However, by using the proposed methodology in this paper, one does not need a different multilevel approximation and framework for each individual scoring measure; instead any standard verification technique, such as the calibration or continuous ranked probability score (CRPS) \citep{Gneiting07}, can be employed on this standard single ensemble forecast.

To generate this ensemble forecast, inverse transform sampling is used. The new single ensemble forecast preserves the unbiased approximation to the mean of the forecast distribution from the original multilevel estimator and forms consistent approximations to other statistics, such as higher moments.

This study proceeds as follows; an introduction to MLMC will be given in Section 2, then a simple method to find a consistent ensemble forecast from a MLMC approximation will be given in Section 3 alongside a corresponding verification technique for these ensemble forecasts. Finally, a conclusion follows.

\section{Multilevel Monte Carlo}

Multilevel Monte Carlo \citep{Giles} is primarily used as a computationally cheap alternative to an equivalent accuracy single level Monte Carlo estimator of statistics with respect to a probability distribution. Suppose one wishes to compute estimates to statistics of $f(X_{L,t})$, such as $\mathbb{E}[f(X_{L,t})]$,
where $X_{L,t}$ is a numerical approximation of our `forecast' random variable $X$
(with discretization parameter $h_{L} \propto M^{-L}$, $M>1$) at time $t \geq 0$ and $f$ is some scalar observable function. Let $X^{i}_{L,t}$,
$i=1,\ldots,N$, be $N \geq 1$ i.i.d. samples of the random variable
$X_{L,t}$. Then an empirical approximation to the density of $X_{L,t}$ is 
\begin{equation}
\pi_{L,t}^{MC}(x)=\frac{1}{N}\sum^{N}_{i=1}\delta(x-X_{L,t}^{i}),
\label{equation:PDF}
\end{equation}
where $\delta$ is the Dirac delta function. One can then estimate statistics to this empirical distribution via the Monte Carlo method. For example, the standard estimator for $\mathbb{E}[f(X_{L,t})]$ is given by
\begin{equation}
\bar{f}_{L,t}^{MC}=\frac{1}{N}\sum^{N}_{i=1}f(X_{L,t}^{i}).
\label{equation:MonteCarlo}
\end{equation}
Now consider the multilevel framework, using $L+1$ ensembles $\{X_{l-1,t}^{i},X_{l,t}^{i}\}_{l=0,i=1}^{l=L,i=N_l}$ (with $X_{-1}^{i}=0$) of sizes $N_{l}$, to derive the equivalent MLMC approximation to $\mathbb{E}[f(X_{L,t})]$,
\begin{equation}
\bar{f}_{L,t}=\frac{1}{N_{0}}\sum^{N_{0}}_{i=1}f(X_{0,t}^{i})+\sum^{L}_{l=1}\Big(\frac{1}{N_{l}}\sum^{N_{l}}_{i=1}(f(X_{l,t}^{i})-f(X_{l-1,t}^{i}))\Big).
\label{equation:MLestimator}
\end{equation}
Taking the telescoping sum of expectations,
\begin{equation}
\mathbb{E}[f(X_{L,t})]=\mathbb{E}[f(X_{0,t})]+\sum^{L}_{l=1}\mathbb{E}[f(X_{l,t})]-\mathbb{E}[f(X_{l-1,t})],
\label{equation:linearityexpectation}
\end{equation}
and considering
\begin{equation}
\mathbb{E}[\hat{f}_{l,t}]=
\begin{cases}
\mathbb{E}[f(X_{0,t})],& l=0, \\
\mathbb{E}[f(X_{l,t})]-\mathbb{E}[f(X_{l-1,t})],& l>0 ,
\end{cases}
\label{equation:unbiased}
\end{equation}
where
\begin{equation}
\hat{f}_{l,t}=
\begin{cases}
\sum^{N_{0}}_{i=1}\frac{1}{N_{0}}f(X_{0,t}^{i}),& l=0, \\
\sum^{N_{l}}_{i=1}\frac{1}{N_{l}}\big(f(X_{l,t}^{i})-f(X_{l-1,t}^{i})\big),& l>0,
\end{cases}
\label{equation:MLMC}
\end{equation}
one recovers $\bar{f}_{L,t}$ as an unbiased approximation of $\mathbb{E}[f(X_{L,t})]$. 
The important thing to note here is that the fine (level $l$) and coarse
(level $l-1$) samples in 
the difference estimators, $\hat{f}_{l,t}$, must be positively
correlated for each $i$. This can be achieved by using the same random system input
(e.g. initial conditions/stochastic forcing) for each $i$ on both
levels. On the other hand, the samples in different ensembles must be uncorrelated. 
The uses of the above framework are incredibly varied. One can even condition these multilevel estimators on observations using processes such as filtering \citep{Jasra, Gregory, Gregory16}. In addition to this, there have been many other applications of MLMC, some of which are highlighted in the review \cite{GilesReview}.

Given an optimal choice of $L$ and $N_{l}$, one can compute these estimators, with the same accuracy as their standard Monte Carlo counterparts, for significantly less computational expense. This works by noting that due to the correlation between the pairs of
samples in each difference estimator, the sample variance of $f(X_{l,t})-f(X_{l-1,t})$, denoted $V_{l}$, should decrease asymptotically with $l \to \infty$. If one desires the accuracy of
$\bar{f}_{L,t}$ to be
\begin{equation}
\mathbb{E}\left(\left(\bar{f}_{L,t}-\mathbb{E}[f_{L,t}]\right)^{2}\right) < \epsilon^{2},
\end{equation}
then one can follow the algorithm in \cite{Giles} to compute $\bar{f}_{L,t}$ by updating, on-line
(as you add additional samples),
the optimal sample sizes
\begin{equation}
N_{l}=\ceil*{2\epsilon^{-2}\left(V_{l}h_{l}\right)\left(\sum_{n=0}^{L}\sqrt{V_{n}/h_{n}}\right)},
\end{equation}
whilst increasing $L$ until $\abs*{\hat{f}_{L,t}}<\frac{1}{\sqrt{2}}(M-1)\epsilon$. An estimated $V_{l}$ can be used in the optimal sample size formula.
Computational cost reductions occur because, if $V_{l}$ decreases asymptotically with $l \to \infty$, then $N_{l}$ also does, leading to a trade-off between estimator variance and bias in each difference estimator. To conclude, we should have large ensembles for the lower levels, and smaller ensembles on the higher levels, given by asymptotically decreasing values of $V_{l}$.


For the full algorithm and corresponding theory, see \cite{Giles, GilesReview}.

\section{Ensemble Forecasting}

This paper now proposes a method to generate a single ensemble forecast from the hierarchy of ensembles created from the MLMC method. Put simply, one can generate a large ensemble (much larger than the finest level ensemble) that represents the entire MLMC approximation to the forecast distribution. This is more useful for the verification of the hierarchy of ensembles rather than simply using standard verification techniques on the finest ensemble in this hierarchy. As mentioned in the previous section, the sample sizes, $N_{l}$, for the pairs of ensembles
on all levels decrease asymptotically, and thus the finest ensemble is the smallest ensemble in the hierarchy. Using the finest ensemble for the verification of the entire MLMC approximation of the forecast distribution would neglect the majority of samples, on lower levels, from which the approximation was composed.

In addition to this verification, given the statistical consistency of this ensemble forecast with the multilevel ensemble hierarchy, many statistics can be easily estimated via this ensemble. 

\subsection{Multilevel Monte Carlo Ensemble Forecasts}

Now assume $X_{l,t} \in \mathbb{R}$, and so if $X_{l,t}$ was multivariate in the section before, $X_{l,t}$ now represents $f(X_{l,t})$ , the scalar observable, such as $X_{l,t}$ evaluated at a point in space. Here we describe how to generate the single ensemble forecast of a scalar observable $X_{F,t}^{i}\in \mathbb{R}$, $i=1,...,N$ from the MLMC hierarchy of ensembles through inverse transform sampling. It is important to note that this ensemble does not contain i.i.d samples from the forecast distribution, instead they will simply be approximations to these samples. However, this single ensemble has the properties to form a consistent empirical estimate to the forecast distribution and associated distribution functions.

From here onwards, we will assume that values of $N_{l}$ and $L$ have been either set or found, and that the
hierarchy of ensembles $\{X_{l-1,t}^{i},X_{l,t}^{i}\}_{l=0,i=1}^{l=L,i=N_l}$, with $X_{-1,t}=0$, has been
generated.
Predominantly, this is because the framework that this paper presents is designed for evaluating any given MLMC
approximation. Each approximation has a hierarchy of ensembles that use values of $N_{l}$ and $L$ that have been optimised around
minimising the cost of that particular approximation. Each approximation typically comes with it's own algorithm to
set-up these values. Thus, by making the aforementioned assumption we can keep this framework general to all approximations.
In addition to this, it is likely that in real forecasting practice
one would pick the desired maximum level $L$ and then set fixed values of $N_{l}$ based on the maximum computational
expense one can use on a particular level.
This way of choosing $N_{l}$ and $L$ is implemented in the numerical example later in the paper.

Inverse transform sampling is the process of evaluating an (approximation to the) inverse CDF, $F^{-1}(u)$, $u \in [0,1]$, also known as the \textit{quantile function}. In the case where the CDF, $F$, of a random variable is strictly increasing and absolutely continuous, there exists a unique value $x \equiv F^{-1}(u)$ for which $F(x)=u$. This distribution must usually be estimated empirically. If the true CDF of the forecast distribution is known to be absolutely continuous and the samples are sorted to form order statistics, then some of these estimates have been shown to be consistent approximations to $F^{-1}(u)$ \citep{Ma}. A very simple consistent estimate for an evaluation to the quantile function, of the distribution with CDF, $F$, using the (ascending) sorted samples $\big\{X^{i}\big\}_{i=1,...,N} \sim F$, $X^{1} < X^{2} < .... < X^{N}$ is,
\begin{equation}
\hat{F}^{-1}(u)=X^{\ceil{N \times u}}.
\label{equation:SampleQuantileFunction}
\end{equation}
Here, the estimate is a consistent one in the sense that it converges in probability to $F^{-1}(u)$ as $N \to \infty$.
One can use linear interpolation and extrapolation to smooth this consistent estimate. Other inconsistent techniques include fitting a parametric distribution to the ensemble, such as a Gaussian, and sampling from a closed form quantile function (e.g. $\Phi$ for a Gaussian distribution) for that distribution. In all cases, when the empirical quantile function is evaluated with i.i.d uniform samples $u \in [0,1]$, approximations to samples of $X$ can be generated.

The use of inverse transform sampling alongside MLMC was first suggested in \cite{GilesBook}. Here it was proposed to be used to minimise the discrete Wasserstein distance between the two paired ensembles in each difference estimator within (\ref{equation:MLestimator}) and thus positively couple them. Instead, here we will use inverse transform sampling in the context of a MLMC approximation to the quantile function of the forecast distribution,
\begin{equation}
\bar{F}_{L,t}^{-1}(u) = R(X)_{0,t}^{\ceil{N_{0} \times u}}+\sum^{L}_{l=1}\big(R(X)_{l,t}^{\ceil{N_{l} \times u}}-R(X)_{l-1,t}^{\ceil{N_{l} \times u}}\big),
\label{equation:MLestimatorquantile}
\end{equation}
where $R(X)^{i}_{l}$ is the $i'th$ order statistic of $X_{l}$, so that $R(X)^{1}_{l}<R(X)^{2}_{l}<...<R(X)^{N_{l}}_{l}$. 
Note that there is not an exact cancellation in expected values of the above estimator terms, as in
the telescoping sum of expectations in (\ref{equation:linearityexpectation}), as
the individual approximations on each level are not unbiased, only consistent in the limit of $N_{l} \to \infty$.
The following algorithm demonstrates how to generate an ensemble $\big\{X_{F,t}^{i}\big\}_{i=1,...,N}$ of arbitrary size $N$, approximating samples of $X_{L,t}$.

\BEGINALG
\Procedure{}{}
\For{$l=0,...,L$}
\If{$l=0$}
\State Sort $X_{0,t}^{j}$, $j=1,...,N_{0}$, so that $R(X)_{0,t}^{1}<R(X)_{0,t}^{2}<...<R(X)_{0,t}^{N_{0}}$
\Else
\State Sort $X_{l,t}^{j}$, $X_{l-1,t}^{j}$, $j=1,...,N_{l}$, so that $R(X)_{l,t}^{1}<R(X)_{l,t}^{2}<...<R(X)_{l,t}^{N_{l}}$ and $R(X)_{l-1,t}^{1}<R(X)_{l-1,t}^{2}<...<R(X)_{l-1,t}^{N_{l}}$
\EndIf
\EndFor
\For{$i=1,...,N$}
\State Set $X_{F,t}^{i}=0$
\State Sample $u^{i} \sim U[0,1]$
\For{$l=0,...,L$}
\If{$l=0$}
\State $X_{F,t}^{i}+=R(X)_{0,t}^{\ceil{N_{0} \times u^{i}}}$
\Else
\State $X_{F,t}^{i}+=R(X)_{l,t}^{\ceil{N_{l} \times u^{i}}}-R(X)_{l-1,t}^{\ceil{N_{l}\times u^{i}}}$
\EndIf
\EndFor
\EndFor
\EndProcedure
\end{algorithmic}

Note that these $X_{F,t}^{i}$ are not samples from $X_{L,t}$, they are only consistent approximations to the evaluations of $F^{-1}_{L,t}(u)$ for a particular $u$. More specifically, for a random uniform sample $u \sim U[0,1]$, we have
\begin{equation}
x=\bar{F}^{-1}_{L,t}(u),
\end{equation}
and as $N_{l} \to \infty$ for all $l$,
\begin{equation}
x \xrightarrow{p} F^{-1}_{L,t}(u),
\end{equation}
where $N_{l}$ are the number of samples used in each difference estimator in (\ref{equation:MLestimatorquantile}). Then in this limit, $x$ converges in probability to a sample from the forecast distribution on the finest level, i.e. $x \sim X_{L,t}$. Therefore any statistical estimate using these samples is a consistent one within this limit.

The single ensemble $\{X_{F,t}^{i}\}_{i=1,...,N}$ can form valid and consistent approximations to statistics of the forecast distribution. For example, the empirical, consistent, CDF of this ensemble forecast found from the MLMC approximation to the forecast distribution is,
\begin{equation}
\hat{F}_{X_{F,t}}(x)=\frac{1}{N}\sum^{N}_{i=1}\mathbb{I}_{X_{F,t}^{i} \leq x}
\end{equation}
where $\mathbb{I}$ is the indicator function. Clearly this is non-decreasing for continuous $X_{F,t}$ and has the support of $[0,1]$.

One assumes that in practice the computational effort of evaluating the above function a large number of times to generate the ensemble $\big\{X_{F,t}^{i}\big\}_{i=1,...,N}$ is negligible in comparison to the expense of generating the original samples on all of the different levels. Thus, the method seems likely to be admissible even when $N$ is much larger than $N_{0}$. Having said this, it makes sense here to set $N \propto N_{0}$ so that both aspects of the approximation (inverse CDF estimator and the ensemble forecast) converge in probability simultaneously. We take $N=\alpha N_{0}$ with $\alpha \in \mathbb{Z}$, $\alpha \geq 1$ for simplicity.

The proposed ensemble forecast also preserves the unbiasedness of the approximation to the first moment of the forecast distribution from the original MLMC approximation. To show this let $\bar{X}_{F,t}=\frac{1}{\alpha N_{0}}\sum^{\alpha N_{0}}_{i=1} X_{F,t}^{i}$ be the sample mean of the ensemble forecast from the multilevel hierarchy of ensembles. Then,

\begin{equation}
\begin{split}
\bar{X}_{F,t}&=\frac{1}{\alpha N_{0}}\sum^{\alpha N_{0}}_{i=1} X_{F,t}^{i}\\
\quad &=\Big(\frac{1}{\alpha N_{0}}\sum^{\alpha N_{0}}_{i=1}\hat{F}^{-1}_{0,t}(u^{i})\Big) +\\
\qquad &\sum^{L}_{l=1}\bigg(\Big(\frac{1}{\alpha N_{0}}\sum^{\alpha N_{0}}_{i=1}\hat{F}^{-1}_{l,t}(u^{i})\Big) -\Big(\frac{1}{\alpha N_{0}}\sum^{\alpha N_{0}}_{i=1}\hat{F}^{-1}_{l-1,t}(u^{i})\Big)\bigg) , \\
\quad &=\Big(\frac{1}{\alpha N_{0}}\sum^{\alpha N_{0}}_{i=1}X_{0, t}^{\ceil*{N_{0} \times u^{i}}}\Big) +\\
\quad &\sum^{L}_{l=1}\bigg(\frac{1}{\alpha N_{0}}\sum^{\alpha N_{0}}_{i=1}\Big(X_{l, t}^{\ceil*{N_{l} \times u^{i}}} - X_{l-1, t}^{\ceil*{N_{l} \times u^{i}}}\Big)\bigg),
\end{split}
\end{equation}

and given that $u^{i}$ are i.i.d. draws of the uniform distribution $Unif[0,1]$, $i=1,...,\alpha N_{0}$ then

\begin{equation}
\begin{split}
\mathbb{E}[\bar{X}_{F,t}]&=\Big(\frac{1}{\alpha N_{0}}\sum^{\alpha N_{0}}_{i=1} \mathbb{E}[X_{0,t}]\Big) +\\
\qquad &\sum^{L}_{l=1}\bigg(\frac{1}{\alpha N_{0}}\sum^{\alpha N_{0}}_{i=1}\Big(\mathbb{E}[X_{l,t}] - \mathbb{E}[X_{l-1,t}]\Big) \bigg)\\
\quad &=\mathbb{E}[X_{0,t}]+\sum^{L}_{l=1}\mathbb{E}[X_{l,t}-X_{l-1,t}]=\mathbb{E}[X_{L,t}].
\end{split}
\end{equation}

\subsection{Assessing the Calibration of Multilevel Monte Carlo Ensemble Forecasts}

Evaluating the ensembles used in ensemble forecasts is very important in checking the predictive value of the forecast. The remainder of this paper concentrates on a method of evaluating the calibration of forecasts from the MLMC approximations to the forecast distribution, directly via the single ensemble forecast found in the previous section: the Probability Integral Transform Histogram. This technique uses observations from the target distribution to evaluate ensemble forecasts. Calibration is the measure of whether the observations are indistinguisable from the samples of the ensemble forecast distribution \citep{Carney}. This is a quality of the empirical forecast distribution that is possibly disregarded if one were to simply study errors of point statistical estimators.

Consider the target distribution, $Y_{obs,t_{k}}$, behind the observed process, where partial observations $y_{obs,t_{k}}$, are taken from a single realisation of this process at times $t_{k}$, $k \in [0,N_{y}]$, $t_{0}=0$, $t_{N_{y}}=T$. Clearly, our aim would be to use a forecast distribution associated with the random variable $X_{t_{k}}=Y_{obs,t_{k}}$, however in many real-world scenarios, $Y_{obs,t_{k}}$ is unknown. Therefore verification techniques are used to rank forecasts on their similarity to the observed process, with the aim of finding the best forecast / model that derived them. The case of $X_{t_{k}}=Y_{obs,t_{k}}$ is known as the random variable with associated forecast distribution from the perfect model.

\subsubsection{Probability Integral Transform Histogram}

The Probability Integral Transform (PIT) histogram is used to determine the uniformity of the observations with respect to the (empirical) CDF of the ensemble, and thus the calibration of the forecast distribution with respect to the target distribution. One can define a random variable $R \sim F_{L,t_{k}}(Y_{obs,t_{k}})$, the Probability Integral Transform. Then samples of $R$ are given by,
\begin{equation}
r_{t_{k}}=F_{L,t_{k}}(y_{obs,t_{k}}).
\end{equation}
The forecast distribution is said to be calibrated with respect to the target distribution if $R \sim Unif[0,1]$, and so a histogram of $r_{t_{k}}$ would be relatively flat. Using the MLMC approximation to the forecast distribution, define the associated multilevel empirical PIT samples,
\begin{equation}
\hat{r}_{t_{k}}=\frac{1}{N}\sum^{N}_{i=1}\mathbb{I}_{X_{F,t_{k}}^{i} \leq y_{obs,t_{k}}}
\end{equation}
where $X_{F,t_{k}}^{i}$ are an arbitrary $N$ members of the ensemble forecast from the multilevel hierarchy of ensembles at time $t_{k}$ using the aforementioned inverse transform sampling method. This is simply the empirical cumulative distribution function of the $N$ ensemble forecast members $X_{F,t_{k}}^{i}$. Here, given that we set $N \propto N_{0}$, then in the limit of $N_{l} \to \infty$, for all $l=0,...,L$,
\begin{equation}
X_{F,t_{k}} \sim F_{L,t_{k}}^{-1},
\end{equation}
and thus $F_{L,t_{k}}(X_{F,t_{k}}) \sim U[0,1]$. By considering this, we have a consistent estimate to the PIT sample $r_{t_{k}}$, when concentrating on the limit of $N \propto N_{0} \to \infty$. One can find the frequency ($H_{i}$, $i=1,...,B$) of $B$ evenly spaced bins in a histogram of these samples by:
\bigskip
\begin{enumerate}
\item Set $H_{i}=0$, for $i=1,...,B$.
\item For each $k=1,...,N_{y}$, find the $i=1,...,B$ in which $\frac{i-1}{B} \leq \hat{r}_{t_{k}} \leq \frac{i}{B}$ and set $H_{i}=H_{i}+1$.
\end{enumerate}
\bigskip
This histogram will be refered to as the multilevel PIT histogram (MLPIT) for the remainder of this paper. The MLMC approximation that derives the ensemble forecast $\{X_{F,t_{k}}^{i}\}_{i=1,...,N}$ can then be described as calibrated with respect to the target distribution if $H_{i} \approx \frac{N_{y}}{B}$ for each $i=1,...,B$. Thus, this can be used to test the variance and biasedness of the ensembles with respect to the target distribution. If the histogram is convex then the ensembles are said to be overdispersed, whereas if it is concave, then the ensembles are said to be underdispersed, and if it is skewed then there exists a bias in the ensembles \citep{Carney}. This is therefore a very appropriate way to clarify if there is any additional bias from the cancellation of intermediate estimators in a MLMC approximation, thus negating the telescoping sum of expectations in (\ref{equation:linearityexpectation}), although this is not demonstrated in this paper.

\bigskip

\textit{\textbf{Example:}}
The following linear mean reverting OU process, $X_{t} \in \mathbb{R}$,
\begin{equation}
\label{eq:OU}
dX_{t}=\alpha(\mu-X_{t})dt + \sigma^{2}d W_{t} ,
\end{equation}
over time time interval $t \in [0,T]$, where $W_{t}$ is a univariate Brownian Motion, will be used alongside pre-defined scenarios of calibration for a MLMC approximation to the forecast distribution to provide a demonstration of the proposed method. We let the observations come from the above model, discretized with timestep $h=2^{-5}$, with $\alpha=0.1$, $\mu=0$ and $\sigma^{2}=0.1$.
In this example, an Euler-Maruyama numerical scheme will be used to discretize
the OU process.
To frame this problem in a likely forecasting setting, we first choose a fixed finest resolution
that we desire, $L=4$ and so $l \in [0, 4]$. A maximum computational expense that we are allowed to use on propagating the entirity of samples in each level of the ensemble hierarchy, $C_{max}=1.536 \times 10^{7}$, is then set. The cost of each sample in $l$'th difference estimator is $\left(Th_{l}^{-1}(1 + 1/2)\right)$ (as all but the first difference estimators in (\ref{equation:MLestimator}) require
coarse and fine time-steps of the discretization) where $h_{l}=2^{-1-l}$ and so $N_{l}$ is given by
\begin{equation}
\begin{split}
N_{l} &= \floor*{\frac{C_{max}}{T\left(h_{l}^{-1}\left(1 + 1/2\right)\right)}}\\
\quad &= \floor*{\left(\frac{2}{3}\right)C_{max}T^{-1}h_{l}}.
\end{split}
\end{equation}
This corresponds to $N_{0} = 2^{7}$.
The arbitrary number of samples $X_{F}^{i}$ to draw from the MLMC approximation to the inverse CDF is set to $N=8N_{0}=2^{10}$.

Pairs of samples from coarse and fine ensembles in each difference estimator in (\ref{equation:MLestimator}) are positively coupled by using the same underlying Brownian Motion, as in \cite{Giles}.
The models are run over times $t \in [0,40000]$ (the long run time is to give the stationary distributions a chance to be simulated), and observations are collected at $t_{k}=k$, $k \in [1,40000]$. At each of these times, a single ensemble forecast is generated from the hierarchy of ensembles that build up the MLMC approximation to the forecast distribution, and is used to verify the calibration of the approximation. Model parameters for four experimental setups are given as follows: $\alpha = 0.1$, $\sigma^{2}=0.1$, $\mu=0$ for the calibrated scenario, $\alpha = 0.1$, $\sigma^{2}=0.02$, $\mu=0$ for the underdispersed scenario, $\alpha = 0.1$, $\sigma^{2}=0.5$, $\mu=0$ for the overdispersed scenario and $\alpha = 0.4$, $\sigma^{2}=0.1$, $\mu=0.2$ for the biased scenario.

\begin{figure}[h!]
  \centering
  \includegraphics[width=94mm]{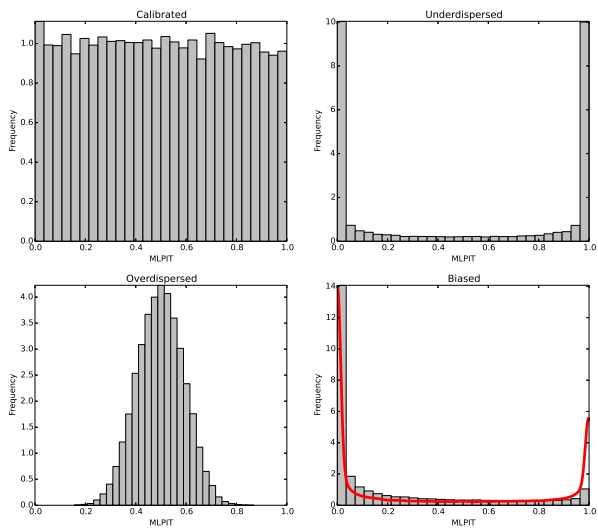}
  \caption{\textit{Multilevel Probability Integral Transform Histograms, using the ensemble $\big\{X_{F}^{i}\big\}_{i=1,...,N}$, of the linear OU process for the four different calibration scenarios. The solid line on the Biased scenario plot shows a smoothed kernel of the PIT histogram generated from the actual stationary forecast and target distributions.}}
  \label{figure:MLPIT_Experiments_LinearSystem}
\end{figure}

\begin{figure}[h!]
  \centering
  \includegraphics[width=94mm]{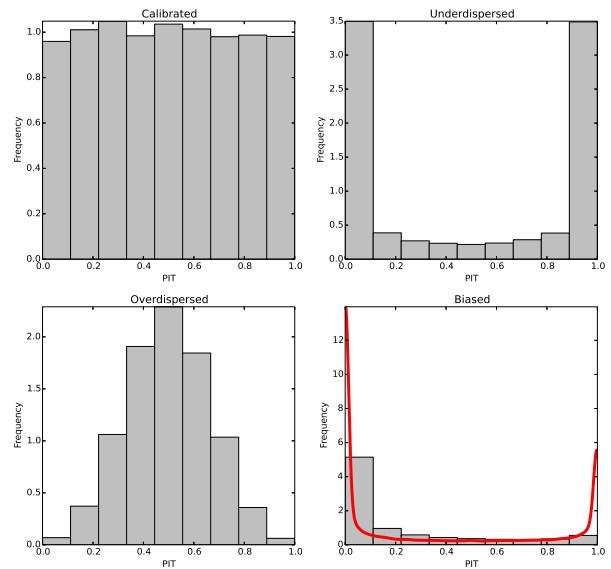}
  \caption{\textit{Probability Integral Transform Histograms, using just the finest ensemble $\big\{X_{L}^{i}\big\}_{i=1,...,N_{L}}$, of the linear OU process for the four different calibration scenarios. The solid line on the Biased scenario plot shows a smoothed kernel of the PIT histogram generated from the actual stationary forecast and target distributions.}}
  \label{figure:PIT_Experiments_LinearSystem}
\end{figure}

This setup allows us to establish that the correct calibration behaviour is being shown by the Multilevel PIT histogram for each of the scenarios, however we will also compare this to the PIT histogram using just the finest ensemble, although this isn't the primary goal of the section. Figures \ref{figure:MLPIT_Experiments_LinearSystem} and \ref{figure:PIT_Experiments_LinearSystem} show the MLPIT and PIT histograms respectively for the four scenarios of calibration listed above. Due to the small number of samples in the finest ensemble, the PIT histogram can only represent a very small number of bins of probability. Both show similar general behaviour for the cases above. 


We can derive the stationary distribution to both the forecast distribution and the target distribution from the model specifications above from the Fokker-Planck equation corresponding to \eqref{eq:OU}. One notes that the stationary forecast distribution using the Biased scenario model above is given by $f \sim N\big(0.2,\frac{1}{8}\big)$ and the stationary target distribution is given by $y \sim N\big(0,\frac{1}{2}\big)$ (as the general form is $\sim N\big(\mu,\frac{\sigma^{2}}{2\alpha}\big)$). Thus the actual PIT histogram can be generated by taking an arbitrarily large number of samples of $F(y)$, where $F$ is the CDF of $f$. A smoothed density kernel of this histogram is superimposed on the corresponding empirical PIT histograms for the single level and multilevel approximations. The empirical histograms approximately match this, however, due to the lack of samples in the finest ensembles, the single level histogram is not as clear to the type or magnitude of bias as shown by the multilevel PIT histogram. This is due to the lack of probability bins in a small, single finest ensemble ($N_{L}+1$), and one would still suffer from similar problems if using interpolation techniques in between the limited number of samples of this ensemble. The MLMC approximations of the forecast distributions and associated histograms are numerically biased (proportional to the finest timestep) from this exact PIT histogram due to the use of a numerical discretization, and so are expected to be slightly different. Despite this, one can clearly interpret the calibration and identify the extent and type of such bias in the MLMC approximations to forecast distributions with more clarity using the multilevel PIT histogram technique proposed here than using standard methods with the small finest ensemble.

\section{Conclusion and Outlook}

This work has discussed the benefits of generating an ensemble forecast from Multilevel Monte Carlo (MLMC) approximations to statistics of random variables representing forecast distributions. The proposed procedure to do this is simple and easily implemented. The calibration of this ensemble forecast has also been examined. Ensemble forecasts provide a simple methodology of deriving empirical estimates to associated distribution functions.  The ensemble hierarchy that forms the computationally efficient MLMC approximations to an arbitrary statistic of the forecast distribution is assumed to have already been generated in preparation for forecasting. It is anticipated that in real forecasting practice, this hierarchy of ensembles would simply be generated by using the maximum ensemble sizes affordable at each level of resolution. The ensemble forecast calibration verification technique takes the entire multilevel hierarchy into account when using the proposed methodology.

Calibration of this ensemble forecast is assessed using the Probability Integral Transform histogram after this single ensemble is generated from the ensemble hierarchy. Thus we have stated what it means for a MLMC approximation to be calibrated with respect to a target distribution. This can be used to evaluate many properties of a MLMC approximation to a forecast distribution including biases (and their type) from intermediate terms in the MLMC telescoping sum of estimators, variances of these approximation and even potentially distribution multimodal feature detection. 

\ack Alastair Gregory was supported by the Science and Solutions to a Changing Planet DTP and the Natural Environment Research Council. He was also supported by the Mathematics of Planet Earth CDT. The authors of this paper would like to thank Chris Ferro (Exeter) for the range of helpful conversations and advice he gave in the development of this research. Thanks also goes to the reviewers of this paper for their constructive feedback on the manuscript.

\bibliography{refs}
\end{document}